\documentstyle[preprint,aps]{revtex}
\tightenlines

\begin{document}
\newcommand{\beq}{\begin{equation}}
\newcommand{\eeq}{\end{equation}}
\newcommand{\beqa}{\begin{eqnarray}}
\newcommand{\eeqa}{\end{eqnarray}}
\newcommand{\fr}{\frac}
\draft
\preprint{INJE-TP-03-02, hep-th/0302148}

\title{No absorption in de Sitter space}

\author{ Y.S. Myung\footnote{Email-address :
ysmyung@physics.inje.ac.kr} and H. W. Lee\footnote{Email-address :
hwlee@physics.inje.ac.kr}}
\address{Relativity Research Center and School of Computer Aided Science,
Inje University, Gimhae 621-749, Korea}

\maketitle

\begin{abstract}
We study the wave equation for  a minimally coupled massive scalar
 in D-dimensional
de Sitter space. We compute the absorption cross section
 to investigate  its cosmological horizon in the southern diamond.
By analogy of the quantum mechanics, it is found that there is no
absorption in  de Sitter space.
This means that de Sitter space is usually  in thermal
equilibrium, like the black hole in anti de Sitter space.
It confirms that the cosmological horizon not
only emits radiation but also absorbs that previously emitted by
itself at the same rate, keeping the curvature radius of de Sitter space
fixed.
\end{abstract}
\vfill
Compiled at \today : \number \time.

\newpage

\section{Introduction}
Recently an accelerating universe has proposed to be a way
to interpret the astronomical data of supernova\cite{Per,CDS,Gar}.
The inflation is employed to solve the cosmological flatness and
horizon puzzles arisen in the standard cosmology.
Combining this observation with the need of inflation
 leads to that our universe approaches de Sitter
geometries in both the infinite past and the infinite future\cite{Wit,HKS,FKMP}. Hence it is
very important to study the nature of de Sitter (dS) space and the
dS/CFT correspondence\cite{BOU,STR}.
 However,
there exist  difficulties in studying de Sitter space.
First there is no spatial infinity and   global timelike
Killing vector.  Thus it is not easy to define  conserved  quantities including  mass,
charge and angular momentum appeared in asymptotically  de Sitter space.
Second the dS solution is absent from string theories and thus we
do not   have a definite example  to test the dS/CFT correspondence.
Third it is hard to define  the $S$-matrix because of the
presence of the cosmological horizon.

We remind the reader that the cosmological horizon is very similar
to the event horizon in the sense that one can define its
thermodynamic quantities of a temperature and an entropy using the same way as was done for
the black hole. Two important quantities in studying the black hole
are the Bekenstein-Hawking entropy and the absorption cross
section (=greybody factor). The former relates to the intrinsic
property of the black hole itself, while the latter relates to the
effect of spacetime curvature. Explicitly the greybody factor for
the black hole arises as a consequence of scattering off the
gravitational potential surrounding the horizon\cite{grey1}. For example,
the low-energy $s$-wave greybody factor for a massless scalar has a
universality such that it
is equal to the area of the horizon for all spherically symmetric
 black holes\cite{grey2}. Also the greybody factor measures the Hawking radiation
 in a semiclassical way.    The entropy for the cosmological horizon was
 discussed in literature\cite{entropy}. However, there is a few
 attempts to compute the greybody factor for the
 cosmological horizon\footnote{A similar work for four-dimensional
 Schwarzschild-de Sitter black hole appeared in\cite{KOY}. But it considered
 mainly the black hole temperature. Also the absorption rate for the Kerr-de Sitter
 black hole was discussed in\cite{STU}. Recently, temperature and entropy of
 the Schwarzschild-de Sitter black hole were discussed in\cite{Shan}}\cite{deabs}.

In this paper we compute the absorption cross section of a massive
scalar in the background of  D-dimensional de Sitter space.
For this purpose we first note that  the wave equation is well defined only in the southern
diamond. Also we should point out a crucial difference between the
cosmological horizon in de Sitter space and the event horizon in
the black hole which states  that de Sitter space is usually assumed to
be in thermal equilibrium\cite{DDO}. This implies that the cosmological
horizon not only emits radiation, but also absorbs radiation
previously emitted by itself at the same rate, keeping the
curvature radius $l$ fixed. On the other hand, one  choose either
a black hole in thermal equilibrium with a heat bath within a bounded box or a black
hole that is truly evaporating. This arises  because the black hole
(radiation in a bounded box) have a negative (positive) specific heat, whereas the
cosmological horizon has a positive specific heat. Two of black
hole and heat bath will be in thermal equilibrium if the box is
bounded. An example is an eternal  black hole in anti de Sitter space
(AdS-black hole)\cite{Mal} because
anti de Sitter space is considered as a box. If the box is unbounded the black hole evaporates
completely, as the Schwarzschild black hole evaporates.
Actually the de Sitter horizon is very similar to the AdS-black hole\cite{DKS}.
In the previous works\cite{deabs}, we did not consider
this stable nature of the cosmological horizon seriously.

We wish to calculate the outgoing flux near $r=0$. And then we
compute the outgoing flux   by
using the matching  region of overlapping  validity  near the cosmological horizon
of $r_c=1$. In this work we will not follow the conventional approach for a computation of the
greybody factor of the black hole. As an analog situation we introduce the
 wave propagation under the potential step with $0<E<V_0$ in
the quantum mechanics. This gives rise to the classical picture of
what the particle goes on :  the total reflection occurs  due to
the potential step. The similar situations also occur in the
scalar wave propagation in the background of the cosmological
horizon. Even though the cosmological
horizon emits radiation, it absorbs radiation
previously emitted by itself at the same rate. Hence there is no
net absorption in a finite time and thus one gets the zero
absorption cross section in the semiclassical approach.

The organization of this paper is as follows. In section II we
briefly review the wave equation in de Sitter space. We perform
a potential analysis to study  the asymptotic region by
introducing a tortoise coordinate $r^*$ in section III.
Also we briefly review the scattering in the potential step of the
quantum mechanics.
In section IV we calculate the  flux at $r=0,1$ to find the greybody
factor.
Finally we discuss our results in section V.

\section{ wave equation in de Sitter space}

We start with the wave equation for a massive scalar
\beq
(\nabla^2 -m^2) \Phi=0
\label{2eq1}
\eeq
in the background of D-dimensional de Sitter space expressed in the static
coordinates
\beq
ds_{dS}^2=-\Big(1- \fr{r^2}{l^2} \Big) dt^2 +
\Big(1- \fr{r^2}{l^2} \Big)^{-1}dr^2 +r^2
d\Omega_{D-2}^2.
\label{2eq2}
\eeq
Here  $l^2$
is the curvature radius of de Sitter space and
hereafter we set $l=1$ for simplicity unless otherwise stated.
$d\Omega^2_{D-2}$ is the metric of the unit sphere $S^{D-2}$.
The above metric is singular at the cosmological horizon, which
divides space into four regions. There are two regions with $0\le r \le1$
which correspond to the causal diamonds of observers at the north
and south poles : northern diamond (ND) and southern diamond (SD).
An observer at $r=0$ is surrounded by a cosmological horizon at
$r=1$.
Two regions with $1<r<\infty$ containing the future-null infinity ${\cal I}^+$ and
past-null infinity  ${\cal I}^-$ are called future triangle (FT)
and past triangle (PT), respectively.
A timelike Killing vector $\fr{\partial}{\partial t}$ is
future-directed only in the southern diamond. To obtain the
greybody factor, we have to obtain a definite wave propagation
as  time evolves. Hence in this work we confine ourselves to the southern diamond.
This means that our working space is compact, in contrast to the
case of the black hole. This gave rise to an ambiguity to
derive the greybody factor in the previous approach\cite{deabs}.

In connection with the dS/CFT correspondence,
one may classify the mass-squared $m^2$ into three cases\cite{STR} :
$m^2\ge1,~~0<m^2<1,~~m^2=0$.
For a massive scalar with  $m^2\ge1$, one has a non-unitary CFT~\footnote{
Although this belongs to one of examples of non-unitary theories
that are dual to well-behaved stable
bulk theory, but the connection between the bulk theory and its non-unitary boundary theory
is not understood clearly  up to now.}. A scalar with mass $0<m^2<1$
can be related to a unitary  CFT.  A massless scalar with $m^2=0$
is special and it would be treated separately.
For our purpose we consider $m^2$ as a parameter at the beginning.
Assuming a mode solution
\beq
\Phi(r,t,\Omega)=f_{\ell}(r) e^{-i \omega t} Y^m_{\ell}(\Omega),
\label{2eq3}
\eeq
Eq.(\ref{2eq1}) leads to the differential equation for
$r$\cite{BMS,ACL}
\beq
(1-r^2)f_{\ell}''(r) + \Big( \fr{D-2}
{r} -Dr \Big) f_{\ell}'(r) + \Big( \fr{\omega^2}{1-r^2}
-\fr{\ell(\ell +D-3)}{r^2} -m^2 \Big) f_{\ell}(r)=0,
\label{2eq4}
\eeq
where the prime ($'$) denotes the differentiation with respect to
its argument. Here $Y_{\ell}^m(\Omega)$ are the hyper-spherical
harmonics on $S^{D-2}$ with
$\nabla^2_{D-2}Y_{\ell}^m(\Omega)=-\ell(\ell+D-3)Y_{\ell}^m(\Omega)$.
In $Y_{\ell}^m(\Omega)$, $\ell$ is a positive integer including 0
and $m$ is a collective index of $(m_1,m_2, \cdots,m_{D-3}).$

\section{potential analysis}

We observe from Eq.(\ref{2eq4}) that
 it is not easy to find how scalar waves propagate in the southern
diamond. In order to do that, we must transform the wave equation
into the Schr\"odinger-like equation using a tortoise
coordinate $r^*$\cite{ML}.  Then we can get
wave forms in asymptotic regions of $r^*\to \pm \infty$ through a potential analysis.
We introduce  $r^*=g(r)$ with $ g'(r)=1/r^{D-2}(1-r^2)$ to transform  Eq.(\ref{2eq4})
into the Schr\"odinger-like equation with the energy $E=\omega^2$

\beq
-\fr{d^2}{d r^{*2}} f_{\ell} + V_D(r)f_{\ell}=  E f_{\ell}
\label{3eq1}
\eeq
with a D-dimensional potential
\beq
V_D(r)= \omega^2 + r^{2(D-2)}(1-r^2) \Big[ m^2 + \fr{\ell(\ell+D-3)}{r^2}
 -\fr{\omega^2}{1-r^2}
\Big].
\label{3eq2}
\eeq
Considering $r^*=g(r)= \int g'(r) dr$, one finds for three-dimensional (D=3) de Sitter  space
\beq
r^*= \ln r - \fr{1}{2} \ln\Big[(1+r)(1-r)\Big],~~ e^{2r^*} =\fr{r^2}{1-r^2},~~
r^2=\fr{e^{2r^*}}{1+e^{2r^*}}.
\label{3eq3}
\eeq
For D=4 de Sitter space, one has
\beq
r^*= -\fr{1}{ r} +\fr{1}{2} \ln\Big[\fr{1+r}{1-r}\Big].
\label{3eq4}
\eeq
The explicit form of $r^*(r)$ depends on the spacetime dimension D.
From the above two expressions we confirm that $r^*$ is a tortoise
coordinate such that $r^* \to -\infty (r \to 0)$, whereas $r^* \to \infty (r \to
1)$. We can express the  potential as a function of $r^*$
explicitly only for D=3 case
\beq
V_3(r^*) = \omega^2 + \fr{e^{2r^*}}{(1+e^{2r^*})^2} \Big[ m^2 +
\fr{1+ e^{2r^*}}{e^{2 r^*}} \ell^2 - (1 + e^{2r^*}) \omega^2 \Big].
\label{3eq5}
\eeq
For $D=3,m^2=1,\ell=0,\omega=0.1$,
the shape of this takes a  potential barrier ($\frown$) located at
$r^*=0$. On the other hand, for all non-zero $\ell$,  one finds the
potential step ($\lnot$) with its height $\omega^2+\ell^2$ on the left-hand side of $r^*=0$.
 $V_3(r^*)$ decreases exponentially to zero as $r^*$
increases on the right-hand side.

From the quantum mechanics we conjecture that  traveling  waves appear near
the cosmological horizon of $r^*=\infty$.
But near the coordinate origin of $r^*=-\infty~(r=0)$,
it is not easy to develop a  genuine traveling wave.
Near $r=0~(r^*=-\infty)$ one finds the equation
\beq
\fr{d^2}{d r^{*2}} f_{\ell,-\infty} -\ell(\ell+D-3)r^{2(D-3)} f_{\ell,-\infty}=0.
\label{3eq6}
\eeq
For D=3case, this  gives us a  solution
\beq
f_{\ell,-\infty}(r^*) =A_3 e^{\ell r^*} + B_3 e^{-\ell r^*}
\label{3eq7}
\eeq
which is equivalently rewritten by making use of Eq.(\ref{3eq3}) as
\beq
f_{\ell,r=0}(r) =A_3 r^{\ell} + \fr{B_3}{ r^{\ell}}.
\label{3eq8}
\eeq
For D$\ge4$ de Sitter space, one obtains
\beq
f_{\ell,r=0}(r) =A_D r^{\ell} + \fr{B_D}{ r^{\ell+D-3}}.
\label{3eq9}
\eeq
In the above two equations,
the first terms correspond to  normalizable modes at $r=0~(r^*=-\infty)$, while the
second terms are  non-normalizable, singular modes. As one discards the second
terms
in Eqs.(\ref{3eq7}) and (\ref{3eq9})
for  calculating the Bogoliubov transformation~\cite{BMS}, the first
terms
are only needed  for our purpose. Hence we set $B_D=0$ for D$\ge3$.

On the other hand,
near the cosmological horizon $r_c=1(r^*=\infty)$ one obtains a differential equation
which is irrespective of $D,\ell$
\beq
\fr{d^2}{d r^{*2}} f_{\infty} +\omega^2 f_{\infty}=0.
\label{3eq10}
\eeq
This  has  a  solution
\beq
f_{\infty}(r^*) =C_D e^{-i \omega r^*} + E_D e^{i \omega r^*}.
\label{3eq11}
\eeq
For D=3 de Sitter space, it is equivalently rewritten  as
\beq
f_{r=1}(r) =C_3(1- r^2)^{\fr{i\omega}{2}} + E_3(1- r^2)^{-\fr{i\omega}{2}} .
\label{3eq12}
\eeq
The first wave /second wave in Eq.(\ref{3eq12}) together with $e^{-i \omega t}$
imply  the ingoing $(\gets)$/outgoing $(\to)$ waves across the cosmological
horizon. This picture is based on the observer confined in the
southern diamond.
In order to obtain a connection between $A_3$ and $C_3,~E_3$, let us
consider the case of $\ell^2 > m^2$   with $V_3(r^*)\approx  V_0= \omega^2+\ell^2$
for $ -\infty < r^* \le 0$ and $V_3(r^*) \approx 0$ for $ 0\le r^* < \infty$ with the energy
$E=\omega^2 <V_0$. It corresponds to the problem for a
 potential step $V_0$ with $0<E<V_0$ in the quantum mechanics\cite{quan}. Requiring
the conservation of flux at $r^*=0$ leads to the asymptotic relations
: $C_3 \approx (\ell-i \omega)A_3/(-2i \omega),
~E_3 \approx (\ell+i \omega)A_3/(2i \omega)=C^*_3$.
It seems that  the flux on the left hand side of $r^*=0$ is zero due to
the real function of Eq.(\ref{3eq8}), but the flux of the right hand side
is not zero because of the traveling wave nature of Eq.(\ref{3eq11}).
For D=4 de Sitter space,  from Eqs.(\ref{3eq4}) and (\ref{3eq11})
we have
\beq
f_{r=1}(r) =C_4(1- r)^{\fr{i\omega}{2}} + E_4(1- r)^{-\fr{i\omega}{2}} .
\label{3eq13}
\eeq
We note that the explicit forms of $f_{r=1}(r)$ for D$>$4 de Sitter space depend
on their spacetime dimensions. However, their asymptotic forms will
take the same form as in Eq.(\ref{3eq13}).

Up to now we obtain  asymptotic forms of a scalar wave which
propagates in the southern diamond of de Sitter space.
In order to calculate the absorption cross section, we need to know
an explicit form of wave  propagation in $0 \le r \le 1$. This can be  achieved
only when
solving the differential equation (\ref{2eq4}) explicitly.

Before we proceed, we wish to review further   the scattering in the
potential step for D=3 de Sitter space. To interpret our solution
Eq.(\ref{3eq11}), it is convenient to multiply it by the $1/C_3$
to give a nice form\cite{quan}
\beq
\fr{f_{\infty}(r^*)}{C_3} = e^{-i \omega r^*} + \fr{E_3}{C_3} e^{i \omega r^*}
\approx
e^{-i \omega r^*} + \fr{C_3^*}{C_3} e^{i \omega r^*}.
\label{3eq14}
\eeq
Because of $|C_3^*/C_3|=1$, two waves (the first and second in Eq.(\ref{3eq14}))
 have amplitudes of the same magnitude. As we will see alter, the absolute square of the
 amplitude of a wave must somewhat be proportional to the flux of
 particle. Hence we conclude that the wave function of Eq.(\ref{3eq14})
 describes the situation in which an ingoing wave is reflected
 back to an outgoing wave by the potential step. This
 interpretation is in accordance with the classical picture of
 what the particle goes on. The wave function $f_{\ell,-\infty}(r^*)/C_3$
 in Eq.(\ref{3eq7}) with $B_3=0$ describes the
 penetration of the Schr\"odinger wave into the classically
 forbidden region of $-\infty<r^* \le 0$. The amplitude of penetrating
 wave decreases exponentially as we go further into the forbidden
 region, and at large distance from the potential barrier the
 amplitude is for all practical purpose zero in accordance with
 the classical picture.

\section{Flux calculation}

In order to solve equation (\ref{2eq4}), we first transform it into
a hypergeometric equation using $z=r^2$. Here  the working space still remains
unchanged as $0 \le z \le 1$ covering  the southern diamond.
This equation  takes a form

\beq
z(1-z)f_{\ell}''(z) -\fr{1}{2}[z(D+1)-(D-1)] f_{\ell}'(z) + \fr{1}{4} \Big( \fr{\omega^2}{1-z}
-\fr{\ell(\ell+D-3)}{z} -m^2 \Big) f_{\ell}(z)=0.
\label{4eq1}
\eeq
Here one finds two poles at $z=0,1 (r=0,1)$ and so  makes a further
transformation  to cancel these by choosing a normalizable
solution at $z=0(r=0)$
\beq
f(z)=z^\alpha (1-z)^\beta w(z),~~
\alpha= \fr{\ell}{2},~~\beta= i\fr{\omega}{2}
\label{4eq2}
\eeq
which is in accordance with Eq.(\ref{3eq9}) with $B_D=0$ and Eqs.(\ref{3eq12})
and (\ref{3eq13})\footnote{In\cite{ACL},
to compute the quasi-normal  mode spectrum of de Sitter space,
the authors considered four cases: i) $\alpha=l/2,\beta=i\omega/2
$ ii) $\alpha=l/2,\beta=-i\omega/2$ iii) $\alpha=-(l+D-3)/2,\beta=i\omega/2$
iv) $\alpha=-(l+D-3)/2,\beta=-i\omega/2$. However,
  i) is relevant to the greybody factor calculation.}
.
Then we obtain a hypergeometric equation
\beq
z(1-z)w''(z) + [c-(a+b+1)z] w'(z) -a b~ w(z) =0
\label{4eq3}
\eeq
where $a,b$ and $c$ are given by
\beq
 a= \fr{1}{2} ( \ell + i \omega +h_+),~~ b= \fr{1}{2} ( \ell + i \omega
 +h_-),~~ c= \ell+\fr{D-1}{2}
 \label{4eq4}
 \eeq
 with
 \beq
 h_{\pm}= \fr{1}{2}\Big[D-1 \pm \sqrt{(D-1)^2-4m^2} \Big].
 \label{4eq5}
 \eeq
 One regular solution near $z=0$ to Eq.(\ref{4eq1}) is given
 by\cite{AS}
 \beq
 f_+(z)
 =A_D z^{\ell/2}(1-z)^{i\omega/2} F(a,b,c;z)
 \label{4eq6}
 \eeq
with an unknown constant $A_D$. For D=odd dimensions,
there is the other solution with a
logarithmic singularity at $z=0$ as $ f_-(z)=
 \tilde A_D z^{\ell/2}(1-z)^{i \omega/2}[F(a,b,c;z) \ln z +\cdots]$.
However, both solutions have vanishing
flux at $z=0$ because the relevant part ($z^{\ell/2}$)
is not complex but real.

Now we are in a position to calculate an  outgoing flux at $z=0(r=0,r^*=-\infty)$
which is defined   as
\beq
{\cal F}(z=0)= 2\fr{2 \pi}{i} [f^*z\partial_z f-f z\partial_z
f^*]|_{z=0}.
\label{4eq7}
\eeq
For any kind of real functions near $z=0(r=0)$ including $f_{\pm}$,
 the outgoing $(\to)$/ingoing $(\gets)$ fluxes are obviously given by
\beq
{\cal F}_{out/in}(z=0)=0.
\label{4eq8}
\eeq
This  means that if a wave form is real near $z=0$, one cannot
find any non-zero flux.
We choose a regular  solution of $f_+(z)$ for
further calculation.
To obtain a flux at the horizon of $z=1(r=1)$, we first  use a formula :
\beqa
&&F(a,b,c;z)=
\fr{\Gamma(c)\Gamma(c-a-b)}{\Gamma(c-a)\Gamma(c-b)}
F(a,b,a+b-c+1;1-z)\\ \nonumber
&& ~~+ \fr{\Gamma(c)\Gamma(a+b-c)}{\Gamma(a)\Gamma(b)}(1-z)^{c-a-b} F(c-a,c-b,-a-b+c+1;1-z).
\label{4eq9}
\eeqa
Using $1-z \approx e^{-2r^*}$ near $z=1$, one finds from Eq.(\ref{4eq6}) the
following form:

\beq
f_{+,0\to1}\equiv f_{in} + f_{out}=
 H_{\omega,\ell} e^{-i \omega r^*} +H_{-\omega,\ell} e^{i \omega
r^*}
\label{4eq10}
\eeq
where
\beq
H_{-\omega,\ell}= H_{\omega,\ell}^*=A_D\alpha_{-\omega,\ell},~~
\alpha_{-\omega,\ell}= \fr{\Gamma(1+\ell)\Gamma(i\omega) 2^{i \omega}}
{\Gamma[(\ell +i \omega +h_+)/2)] \Gamma[(\ell +i \omega +h_-)/2)]}.
\label{4eq11}
\eeq
Then we match Eq.(\ref{3eq11}) with Eq.(\ref{4eq10}) to yield $C_D=H_{\omega,\ell}$
and $E_D=H_{-\omega,\ell}$ near the cosmological horizon.
Finally we calculate its outgoing $(\to)$ flux at $z=1(r^*=\infty)$ as
\beq
{\cal F}_{out}(z=1)= \fr{2 \pi}{i} [f_{out}^*\partial_{r^*} f_{out}-f_{out}
\partial_{r^*}
f_{out}^*]|_{r^*=\infty} = 4 \pi \omega A_D^2
|\alpha_{-\omega,\ell}|^2.
\label{4eq12}
\eeq
On the other hand, the ingoing $(\gets)$ flux is given by
\beq
{\cal F}_{in}(z=1)= \fr{2 \pi}{i} [f_{in}^*\partial_{r^*} f_{in}-f_{in}
\partial_{r^*}
f_{in}^*]|_{r^*=\infty} = 4 \pi \omega A_D^2
|\alpha_{\omega,\ell}|^2.
\label{4eq13}
\eeq
Thus  it equals ${\cal F}_{out}(z=1)$
because the modulus of amplitude
$|\alpha_{\omega,\ell}|$ is equal to $|\alpha_{-\omega,\ell}|$.

\section {absorption cross section}
Up to now we do not
insert the curvature radius $l$ of de Sitter space.
The correct absorption coefficient can be recovered when replacing $\omega(m)$ with
$ \omega l(m l)$.
An absorption coefficient by the cosmological horizon is defined formally by
\beq
{\cal A} = \fr{ {\cal F}_{out}(z=1)}{{\cal F}_{out}(z=0)}.
\label{5eq0}
\eeq
However, one cannot define the
absorption cross section for this case
because we have zero-flux of ${\cal F}_{out}(z=0)=0$ for
de Sitter wave propagation. Instead, we may  follow the black hole approach to obtain
the absorption cross section\cite{deabs}.   If we assume an unknown normalization
${\cal F}_{out}(z=0) =H_D$, then $H_D$ may be determined by referring
the absorption cross section for  the low-energy $s(\ell=0)$-wave.
In this way the absorption cross section in three dimensions may be defined
by
\beq
\sigma_{abs}^{D=3}= \fr{{\cal A}_3}{\omega} =
\Big[\fr{4\pi l A_3^2}{H_3}\Big] |\alpha_{-\omega l,\ell}|^2
\label{5eq1}
\eeq
where
\beq
|\alpha_{-\omega l,\ell}|^2=\fr{|\Gamma(1+\ell)|^2|\Gamma(i\omega l)|^2 }
{|\Gamma[(\ell + i\omega l + \tilde h_+)/2)]|^2
|\Gamma[(\ell + i\omega l + \tilde h_-)/2]|^2}.
\label{5eq2}
\eeq
Here $\tilde h_{\pm}$ is obtained by replacing $m$ by $ml$ at
$h_{\pm}$.

For D=4 de Sitter space, one may have
\beq
\sigma_{abs}^{D=4}= \fr{\pi{\cal A}_4}{\omega^2} =
\Big[\fr{4\pi^2 l A_4^2}{\omega H_4}\Big] |\alpha_{-\omega
l,\ell}|^2.
\label{5eq3}
\eeq
However, this approach based on the calculation of the black hole-greybody factor
 leads to a wrong result  to find the absorption cross
section for the cosmological horizon.
Actually there is no   wave propagating truly along
``$-$"axis-direction. As is shown in Eq.(\ref{4eq13}), the
incident (ingoing) wave is totally reflected to give the reflected (outgoing) wave
by the approximate potential step at $r^*\approx 0$. This is confirmed by the
zero flux of ${\cal F}(r^*=-\infty)=0$.

\section{discussion}
We study the wave equation for  a minimally coupled massive scalar
 in D-dimensional
de Sitter space. We compute the absorption cross section
 to investigate  its cosmological horizon in the southern diamond.
By analogy of the quantum mechanics of the wave scattering under the potential step,
 it is found that there is no
absorption of a scalar wave in  de Sitter space in the
semiclassical approach.
This means that de Sitter space is usually stable and  in thermal
equilibrium, unlike the black hole. The cosmological horizon not
only emits radiation but also absorbs that previously emitted by
itself at the same rate, keeping the curvature radius of de Sitter space
fixed. This can be proved by the relation of ${\cal F}_{out}(z=1)
={\cal F}_{in}(z=1)$ and ${\cal F}_{out/in}(z=0)=0$. This exactly
coincides with the wave propagation of the energy $E$ under the
potential step with $0<E<V_0$ which shows the classical picture of
what the particle goes on. Here we find a nature of the eternal de
Sitter horizon\cite{DKS}, which means that
its cosmological constant $\Lambda_D=(D-2)(D-3)/2l^2$ remains
unchanged, as like the eternal black hole in anti de Sitter
space \cite{Mal}.

\section*{Acknowledgement}
Y.S. was supported in part by  KOSEF, Project
Nos. R01-2000-000-00021-0 and R02-2002-000-00028-0.
H.W. was in part supported by KOSEF, Astrophysical Research Center
for the Structure and Evolution of the Cosmos.

\end{document}